\documentclass[11pt]{article}

\newcommand{\ignore}[1]{ }
				
\newcommand{\BibTeX}{{\rm B\kern-.05em{\sc i\kern-.025em b}\kern-.08em
    T\kern-.1667em\lower.7ex\hbox{E}\kern-.125emX}}

\usepackage{epsfig}

\def\CN{{\cal N}}
\def\CK{{\cal K}}
\def\hub{\mbox{hub}}
\def\auth{\mbox{auth}}
\def\flow{\mbox{flow}}
\def\rel{\mbox{Rel}}
\def\precr{\mbox{Precision-at-$r$}}
\def\seekFlow{\mbox{seekFlow}}
\def\factFlow{\mbox{factFlow}}
\def\maxwt{\mbox{maxwt}}
\def\SeekRel{\mbox{\em SeekRel}}
\def\FactRel{\mbox{\em FactRel}}
\def\SurfRel{\mbox{\em SurfRel}}

\title{Relating Web pages to enable information-gathering tasks}

\author{Amitabha Bagchi\\Dept of Computer Science \& Engg\\Indian Institute of Technology\\New Delhi 110016 \and Garima Lahoti\\
Cazoodle Inc.\\
60 Hazelwood Drive, Suite 122,\\
Champaign, IL 61820-7460
}

\begin{document}

%\begin{bottomstuff}
%Author's address: Amitabha Bagchi, Department of Computer Science and Engineering, Indian Institute of Technology, Hauz Khas, New Delhi 110016, India\newline
%\end{bottomstuff}

\maketitle

\begin{abstract}
We argue that relationships between Web pages are
functions of the user's intent. We identify a class of Web tasks -
information-gathering - that can be facilitated by a search engine
that provides links to pages which are related to the page the user is
currently viewing. We define three kinds of intentional relationships
that correspond to whether the user is a) seeking sources of
information, b) reading pages which provide information, or c) surfing
through pages as part of an extended information-gathering process. We
show that these three relationships can be productively mined using a
combination of textual and link information and provide three scoring
mechanisms that correspond to them: {\em SeekRel}, {\em FactRel} and
{\em SurfRel}. These scoring mechanisms incorporate both textual and
link information. We build a set of capacitated subnetworks - each
corresponding to a particular keyword - that mirror the
interconnection structure of the World Wide Web. The scores are
computed by computing flows on these subnetworks. The capacities of
the links are derived from the {\em hub} and {\em authority} values of
the nodes they connect, following the work of Kleinberg (1998) on
assigning authority to pages in hyperlinked environments. We evaluated
our scoring mechanism by running experiments on four data sets taken
from the Web. We present user evaluations of the relevance of the top
results returned by our scoring mechanisms and compare those to the
top results returned by Google's Similar Pages feature, and the {\em
Companion} algorithm proposed by Dean and Henzinger (1999).
\end{abstract}

\section{Introduction}

The tremendous success of collaborative filtering-based recommendation
systems (see e.g.~\cite{liang:2007}) in online retail settings
(e.g. Netflix) has demonstrated that users welcome guidance while
looking for books to buy or films to rent i.e. where they are not
looking for a product which satisfies a general specification rather
than a specific product. In the enterprise search space, the
increasing importance of faceted search - essentially a method of
providing recommendations to satisfy a user's search needs by creating
multiple taxonomies - pioneered by Endeca~\cite{endeca:2006}, under
the name ``guided navigation'', shows that businesses are recognizing
that they can improve profitability by effectively helping their
employees and customers browse through large databases by providing
search results related to the ones users express a preference for.

But search engines for the World Wide Web have been largely
unsuccessful in providing accurate and helpful recommendations to
their users. Research has found that most Web users are not using
advanced features provided by search engines; it has been shown that
they barely understand what these features
do~\cite{spink:2001}. Further, it has been seen that Web users search
less and browse more~\cite{aula:2005}. And, in fact, the process of
developing expertise in using the Web coincides with an increase in
browsing and decrease in searching~\cite{cothey:2002}.

Despite this bleak scenario it is our contention that search engines
have the resources to effectively provide users with related pages
that can help in information-gathering tasks. Further we believe that
a search engine which can do this will increase its value for its
users, with consequent increases in revenue. There is a growing
understanding in the search domain that user intent is crucial to the
search process~\cite{rose:2006,jansen}. Extending this understanding
to the domain of relationships between pages can make
information-gathering tasks easier.

In other words, to know the relationship between two pages, we must
first know what function these two pages serve for the users who visit
them. It is only by characterizing the task the user is engaged in
that we can offer related pages and hope that these will actually
facilitate the task. In this paper we provide a suite of scoring
mechanisms that relate Web pages: {\em SeekRel}, {\em FactRel} and
{\em SurfRel}. The purpose of these scoring mechanisms is to help
identify pages which may be related to the current page depending on
whether the user is reading pages which provide information, seeking
links to sources of information, or surfing through pages as part of
an extended information-gathering process.

A brief overview of our method is as follows: We compute our scores
using flow calculations in a set of subnetworks of the Web. Each
subnetwork corresponds to a single keyword. The set of subnetworks
used to score a pair of pages is decided by finding the keywords
relevant to the pair of pages being scored. Then the edges of these
networks are capacitated using the hub values~\cite{kleinberg} of
their originating pages. Finally flow is sent along these edges
towards special nodes we call {\em witnesses}. The amount of flow that
can be routed is used as a measure of the relationship.

\noindent{\bf Organization.} In Section~\ref{sec:ig} we focus on
information-gather\-ing tasks and identify how our recommendations can
facilitate them. We survey previous work done in relating Web pages in
Section~\ref{sec:related}. Our specific scoring mechanisms are
described in Sections~\ref{sec:algorithm:graph}
and~\ref{sec:algorithm:flow}. A comparison of our results for a small
toy network with the results produced by two recent proposals {\em
PageSim}~\cite{pagesim} and {\em SimRank}~\cite{simrank} is presented
in Section~\ref{sec:algorithm:discuss}. In
Section~\ref{sec:experiments} we present experiments conducted on real
data taken from the Web. We discuss the results of user surveys that
compared the top results produced by our scoring mechanisms with the
results given by the {\em Companion} algorithm of Dean and
Henzinger~\cite{dean} and Google's Similar Pages feature.  Finally in
Section~\ref{sec:discussion}, we conclude by discussing the merits of
our scheme in comparison to these two proposals and by arguing that
our scheme is a better candidate for inclusion in a search engine than
the {\em Companion} algorithm.

\section{Information-gathering on the Web}
\label{sec:ig}

\begin{figure}
\centering
\epsfig{figure=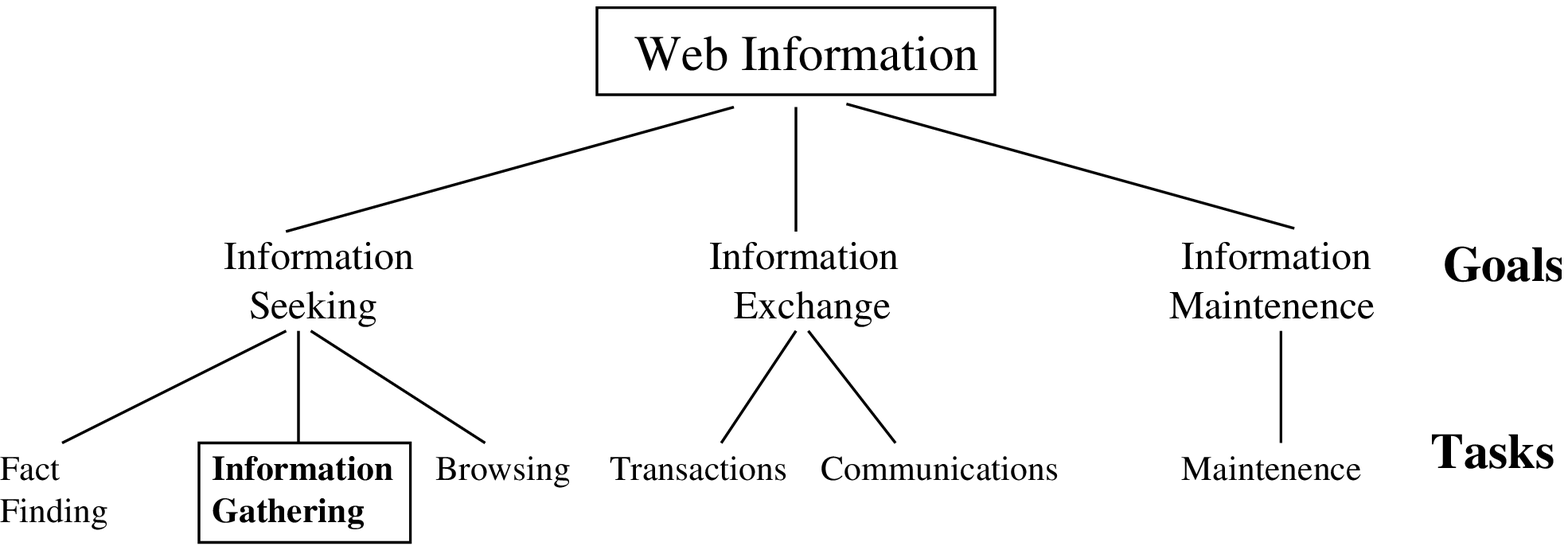,width=0.9\columnwidth}
\caption{Kellar's classification of Web information. Source: Kellar, 2007.}
\label{fig:kellar-classification}
\end{figure}

The tasks that Web users undertake were classified by
Broder~\cite{broder} to generally fall into three categories:
navigational (finding specific pages), informational (seeking facts)
and transactional (performing some interactive set of tasks.)
Kellar~\cite{kellar:2007} further classifies informational tasks into
information-seeking, information exchange and information maintenance
tasks. It is in the first of these classes, information seeking, that
search engines make their major contributions. Kellar differentiates
between three types of information seeking tasks: {\em fact-finding}
e.g. directions to a friend's house or exam dates, {\em
information-gathering} e.g. tectonic movements, Mac laptops, {\em
browsing} news, friend's homepage. See
Figure~\ref{fig:kellar-classification} for a schematic representation
of Kellar's classification (Source~\cite{kellar:2007}.)

Our focus is on information-gathering which we distinguish from fact
finding by using Kellar's definition: {\em Information-Gathering
consists of tasks in which a user is collecting information, often
from multiple sources, in order to write a report, make a decision, or
become more informed about a particular topic}~\cite[pp
67]{kellar:2007}. With this definition in hand let us try to
characterize this class of tasks. 

\subsection{The iterative nature of information-gathering tasks}
\label{sec:ig:characterization}

In their seminal work Belkin, Oddy and
Brooks~\cite{belkin-1:1982,belkin-2:1982} pointed out that assuming
that a user with a need for information will be able to specify that
need exactly is a mistake. Instead, they pointed out that a user comes
to an information retrieval system with an {\em anomalous state of
knowledge}, which she then attempts to express as a query. Based on
the information received the anomaly is rectified to some extent, the
user's image of the world is altered somewhat. But this is not the end
of the process. The altered state of the user's knowledge generates
new anomalies which she then takes back to the information retrieval
engine in the form of requests or queries. This iterative process
continues till the user is satisfied with the extent of the change in
his state of knowledge.

Belkin et. al.'s work has been refined in several directions (we refer
the reader to Marchioni's book for a survey~\cite{marchionini:1995})
but the general view remains a powerful organizing principle for
information retrieval. This general view is borne out in the World
Wide Web setting by Rose's conclusion~\cite{rose:2006} - based on
earlier studies he and Levinson conducted~\cite{rose:2004} - that
information search was an iterative process. Other supporting evidence
in this regard is Aula et. al.'s finding that experienced Web user's
showed a pattern of searching, then browsing, the searching
again~\cite{aula:2005}, which was similar to Cothey's finding in a
longitudinal study that followed subjects evolving from novice to
expert~\cite{cothey:2002}.

\begin{figure}
\centering
\epsfig{figure=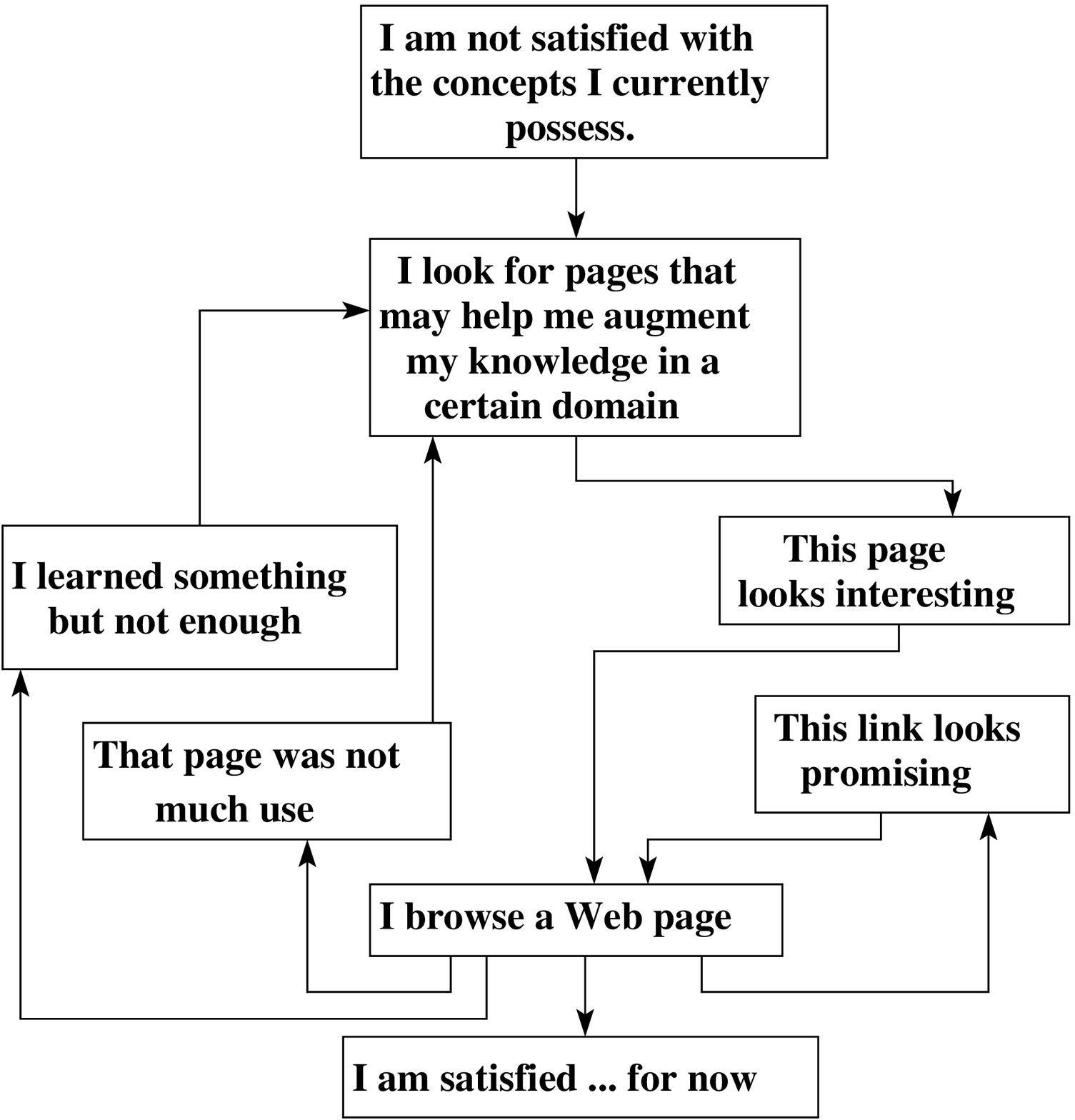,width=0.6\columnwidth}
\caption{A schematic of information-gathering on the Web.}
\label{fig:ig-iterative}
\end{figure}

With all of this as background we characterize the information
gathering task on the World Wide Web as an iterative one (see
Figure~\ref{fig:ig-iterative}). This task is initiated by an
understanding that the user's knowledge needs to be augmented,
proceeds by looking for Web pages (either from search engines or other
sources) and then browsing them when they are found. The process of
browsing rectifies the knowledge anomaly partially or wholly and has
the additional effect of providing links to other pages which might
aid the process. The user may then choose to follow those other links
(immediately or later) or finish browsing the current page. When a
page is browsed completely the user is either satisfied entirely and
terminates the task, or is partially satisfied and resumes the task by
again looking for Web pages.

This iterative characterizations of the information-gathering tasks
suggests ways in which search engines could aid users in performing
them. It is to this that we now turn our attention.

\subsection{Enabling information-gathering tasks}
\label{sec:ig:}

In Figure~\ref{fig:ig-iterative} we notice that a user browsing a Web
page in the process of gathering information treats the page either as
a source of information or as a source of links to other pages. It is
therefore appropriate to provide users with links to two kinds of
pages:
\begin{description}
\item{\bf P1.} Pages which contain information similar to the current
  page.
\item{\bf P2.} Pages which provide links similar to the links provided
  by the current page.
\end{description}

Additionally it is our contention that as user experience with the Web
improves, there will be the realization that people who create Web
content and Web links have an understanding of the interrelationships
between various pages. And so we suggest that a third kind of page
could be useful in the information-gathering process:

\begin{description}
\item{\bf P3.} Pages which can be reached by following a sequence of
  links from the current page and pages from which the current page
  can be reached by following a sequence of links.
\end{description}

The specifications {\bf P1} and {\bf P2} are general and open to
interpretation as to how they might be satisfied. The pages of {\bf
  P3} are more specific but there is still the question of which one
of these pages to choose to present to the user from among the several
thousand that might satisfy this criterion. In
Section~\ref{sec:algorithm} we will provide three scoring
mechanisms: {\em FactRel} will account for pages of type {\bf P1},
{\em SeekRel} for {\bf P2} and {\em SurfRel} for {\bf P3}.

How should these links be provided to the user? While this is a
question better addressed by experts in human-computer interaction, we
would suggest that the ``toolbars'' provided by many search engines
could be augmented to provide page recommendations. These already
provide information about the ``rank'' of the page currently being
viewed and various other pieces of information from the search
engine's bag, and this additional use can fit in seamlessly. For
searches that take place on a search engine's Web page, each ``similar
pages'' link can contain these related pages. In either case it is of
paramount importance that the links provided be appropriately
categorized so that the user can choose to follow them (or not)
depending on the particular function the page being viewed (or the
initial search result) plays in her information-gathering process.

\section{Related work}
\label{sec:related}

The relationship between Web pages has been studied fairly
extensively. Most researchers use the term ``similarity'' and define
it variously. In this paper we have consciously avoided the use of
this term since we believe that the relationship between Web pages is
not intrinsic to the pages but depends on the functionality of the
pages for the user. However, in order to survey previous contributions
to this area, in this section we will use the term ``similarity.''

There are two broad categories of approaches to the Web page
similarity problem. The first relies mainly on the textual content of
Web pages. The pair of pages being compared is either seen just as two
groups of items (keywords, anchor text, patterns in the text) which
overlap significantly or as structured entities that resemble each
other in their organization e.g. which tags appear next to which
ones. Since this is not our approach we do not review this vast
literature here, referring the reader to~\cite{tombros} for a succinct
survey.

The second approach involves taking link and interconnection
information into account. An important way in which interconnection
has been used is to ascribe authority to a page based on which pages
link to it. This idea forms the basis of the Pagerank algorithm
employed by Google~\cite{page}. In a similar vein,
Kleinberg~\cite{kleinberg} described two attributes of Web pages: they
can be {\em authorities} on a topic, or they can be {\em hubs},
linking to pages which are authoritative. He described an iterative
algorithm to compute measures of these two attributes for each Web
page. One line along which Kleinberg's work was developed involved
using anchor text as a descriptive summary of the page being linked
to~\cite{chakrabarti:1998,haveliwala:2002}. But, more relevant to our
methods was the focus on the link structure seen in Dean and
Henzinger's paper~\cite{dean} where they gave two algorithms for
finding similar Web pages. One of their algorithms used the idea of
co-citation earlier seen in Pitkow and Pirolli's
work~\cite{pitkow}. The second algorithm, called {\em Companion}
shares an important feature with our scoring mechanisms: it uses
Kleinberg's algorithm. Their method of building a focused graph
extends Kleinberg's ideas. We have used their ideas for building
capacitated subnetworks in our algorithm. While Dean and Henzinger
give a list of pages similar to a given page, Huang
et. al.~\cite{huang:2004} gave measures of similarity based on the
predecessor and successor set, and also on the basis of all the
vertices reachable from the two pages. Another sort of ``closure'' of
co-citation was used to define a similarity measure called {\em
PageSim} in~\cite{pagesim}. In this measure, pages propagate their
similarity measure to their neighbors, the importance of a particular
propagation being decided by the PageRank of the page. This measure
was an improvement on the {\em SimRank} measure proposed
in~\cite{simrank} which worked on the principle that two pages are
similar if they are linked to by similar pages. The {\em SimRank}
measure was shown to be specific instance of a general framework for
computing similarity between heterogeneous data objects by Xi
et. al.~\cite{xi} who proposed the {\em SimFusion} algorithm. In
Section~\ref{sec:algorithm:discuss} we will compare the results {\em
PageSim} and {\em SimRank} produce on a small toy example with the
results given by our scoring mechanisms.

In a different use of link structure related to our own Lu
et. al.~\cite{lu,lu:2007} claimed that two pages were said to be
similar if flow could be routed from one of them to the
other. However, unlike our work, their capacity assignments were not
based on any notion of authority. To the best of our knowledge this is
the only other mention of using flow to score similarity in the
literature.

As mentioned, the literature on finding similarity is vast and has
seen contributions from many different areas. The papers we have
discussed above are the ones whose techniques are closely related to
our own. With these in view we now proceed to describing our scoring
mechanisms.

\section{Computing relationship scores}
\label{sec:algorithm}

\begin{figure}[htbp]
\begin{center}
\epsfig{file=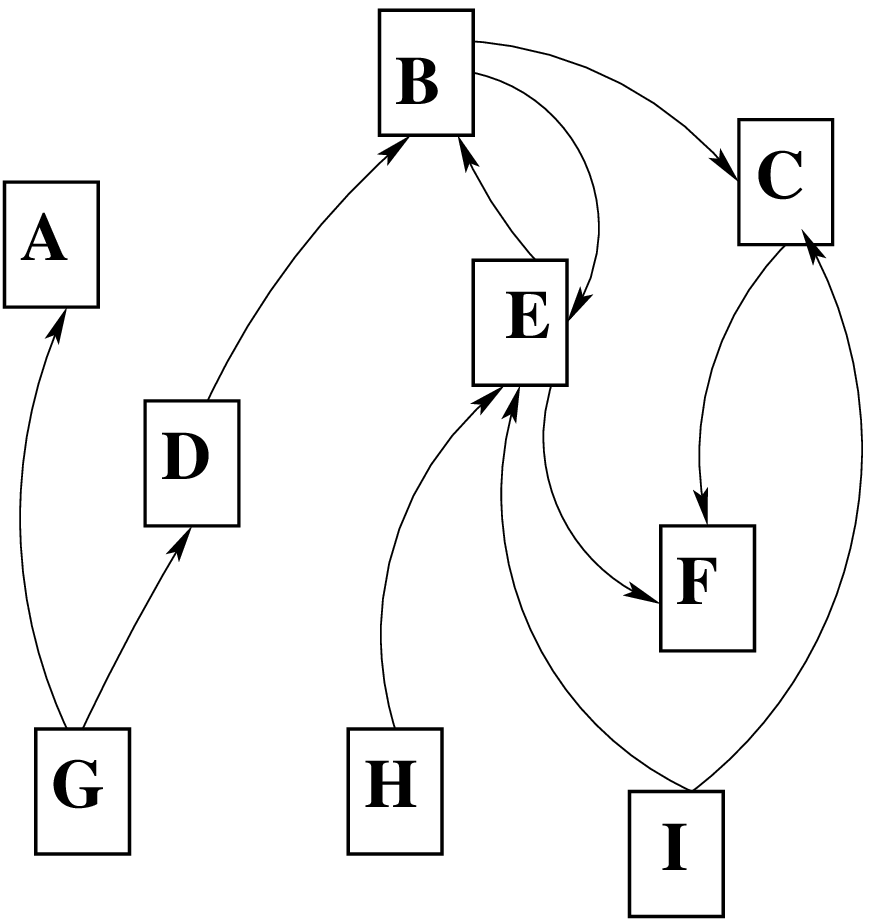,width=0.35\columnwidth}
\caption{A simple hyperlinked network}
\label{fig:witness-example}
\end{center}
\end{figure}

In this section we describe our algorithms for scoring the
relationships between a pair of pages $u$ and $v$. It is our
contention that two pages should have a high {\em FactRel} score if it
is possible to find paths from multiple pages to both of them. We say
that a page $z$ {\em witnesses} the {\em FactRel} relationship between
$u$ and $v$ if it is possible to reach both $u$ and $v$ from $z$ by
following a series of links. For example, in
Figure~\ref{fig:witness-example}, the page $G$ witnesses {\em FactRel}
between $A$ and $B$. Similarly we say that a page $z$ witnesses the
{\em SeekRel} relationship if it is possible to reach $z$ by following
a series of links from $u$ to $z$ and $v$ to $z$. For example, in
Figure~\ref{fig:witness-example}, $C$ witnesses {\em SeekRel} between
$G$ and $I$. $F$ is another such witness for these two. Somewhat
differently, the {\em SurfRel} relationship does not require explicit
witnesses. We say that the {\em SurfRel} relationship exists between
$u$ and $v$ if it is possible to reach $u$ from $v$ by following a
series of links, or vice versa. In Figure~\ref{fig:witness-example},
we see that $H$ and $B$ are related this way, while $A$ and $I$, are
{\em not} related by {\em SurfRel}.

Hence our scores are based on finding paths in the World Wide Web. But
which paths should be given more weight than others in the scoring
process? To answer this question we rely on Kleinberg's
notion~\cite{kleinberg} of {\em authorities} (pages which have
credible information) and {\em hubs} (pages which link to good
authorities) in order to derive a relevant set of focused capacitated
subnetworks from the structure of the World Wide Web. We then
determine relationship scores by sending flow {\em to} the two pages
{\em from}
witnesses (for {\em FactRel}), {\em from} the two pages {\em to} witnesses (for
{\em SeekRel}) or {\em between} the pages (for {\em SurfRel}) in these
capacitated networks.

We will now discuss these methods in detail. In
Section~\ref{sec:algorithm:graph} we will describe how to build our
set of capacitated subnetworks using Kleinberg's method for assigning
authority to Web pages. In Section~\ref{sec:algorithm:flow} we
describe how to find witnesses, route flow and compute scores. We
illustrate the working of the algorithm on a toy example in
Section~\ref{sec:algorithm:discuss} and also present a comparison by
scoring {\em Simrank}~\cite{simrank} and {\em PageSim}~\cite{pagesim}
on the same example.

\subsection{Building capacitated subnetworks}
\label{sec:algorithm:graph}

We begin by assuming that we have a database of significant keywords,
call it $\CK$. We also assume that there is some function $\gamma :
\CK \rightarrow (0,1)$ which assigns relative importance to these
keywords. We do not specify how to build $K$, preferring to rely on
the wealth of tools available for this purpose (see,
e.g.~\cite{yahoo-term-api}). For each of these keywords we apply
Kleinberg's~\cite{kleinberg} HITS algorithm to compute hub and
authority values for all the pages associated with that
keyword. Finally, at the end of this preprocessing step each page has
hub and authority values computed for the subset of the database of
keywords it is associated with. When we are given a pair of pages to
score we identify sets of significant keywords $K_u$ and $K_v$ for $u$
and $v$. We then merge these two sets to get the top $k$ keywords for
some tunable parameter $k$. This set of chosen keywords we call
$K$. Once we have this set of keywords we proceed by making a set of
networks $\CN = \{N_w \mid w \in K\}$. For each $N_w$ we first find
the set of Web pages $P_w$ which contains the keyword $w$.

\begin{figure}[ht]
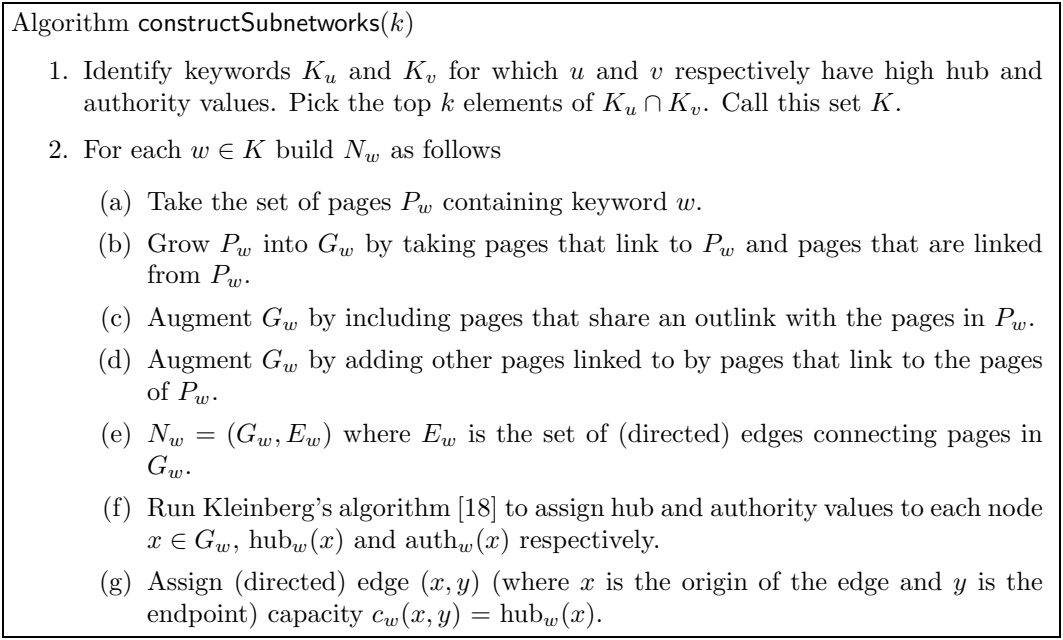

\begin{center}
\fbox{\small
\begin{minipage}{0.8\columnwidth}
\noindent{Algorithm {\sf constructSubnetworks}($k$)}
\begin{enumerate}
\item Identify keywords $K_u$ and $K_v$  for which $u$ and $v$
  respectively have high hub and authority values. Pick the
  top $k$ elements of $K_u \cap K_v$. Call this set $K$.
\item For each $w \in K$ build $N_w$ as follows
\begin{enumerate}
\item Take the set of pages $P_w$ containing keyword $w$.
\item Grow $P_w$ into $G_w$ by taking pages that link to $P_w$ and
  pages that are linked from $P_w$.
\item\label{step:dean} Augment $G_w$ by including pages that
  share an outlink  with the pages in $P_w$.  
\item\label{step:dean2} Augment $G_w$ by adding other pages linked to by
  pages that link to the pages of $P_w$.
\item $N_w = (G_w, E_w)$ where $E_w$ is the set of (directed) edges
  connecting pages in $G_w$.
\item Run Kleinberg's algorithm~\cite{kleinberg} to assign hub and
  authority values to each node $x \in G_w$, $\hub_w(x)$ and
  $\auth_w(x)$ respectively.
\item Assign (directed) edge $(x,y)$ (where $x$ is the origin of the
  edge and $y$ is the endpoint) capacity $c_w(x,y)$ = $\hub_w(x)$. 
\end{enumerate}
\end{enumerate}
\end{minipage}
}
\caption{Building a set of capacitated subnetworks}
\label{fig:algorithm:construct}
\end{center}
\end{figure}

Since we will be using Kleinberg's hub and authority
values~\cite{kleinberg} to capacitate the network (in
Section~\ref{sec:algorithm:flow}) we grow $N_w$ by first taking all
the pages that link to the pages in $P_w$ and the pages that are
linked to by the pages in $P_w$ (as described
in~\cite{kleinberg}). Then we add the refinements to this structure
proposed by Dean and Henzinger~\cite{dean}. Only the main feature of
these refinements has been mentioned in Steps~\ref{step:dean}
and~\ref{step:dean2}. The reader is referred to~\cite{dean} for
further details.

Finally, for each network $N_w$ we run Kleinberg's algorithm for
assigning hub and authority values to each page and label a node
$z$ of network $N_w$ with these values: $\hub_w(z)$ and
$\auth_w(z)$. A directed edge from node $x$ to node $y$ is assigned
capacity $\hub_w(x)$.

A summary of the algorithm is in Figure~\ref{fig:algorithm:construct}.
We postpone a discussion of the motivation for this construction to
Section~\ref{sec:algorithm:discuss}.

\subsection{Using flows to compute relationship scores}
\label{sec:algorithm:flow}

\begin{figure}[ht]
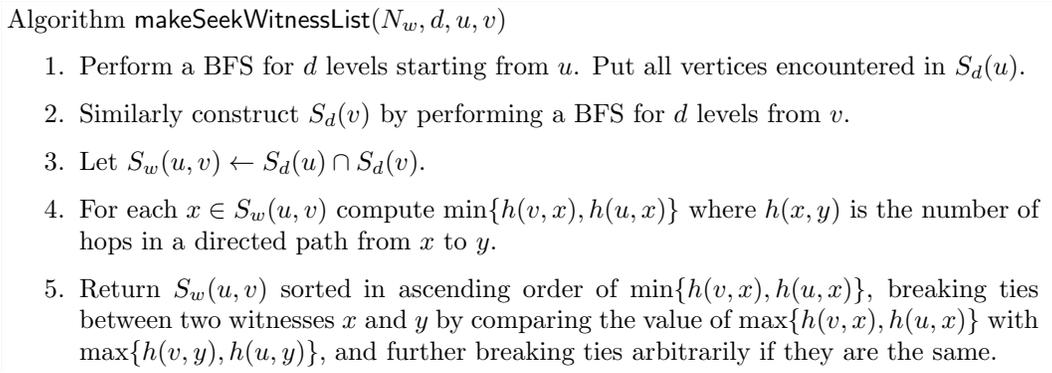

\begin{center}
\fbox{\small
\begin{minipage}{0.8\columnwidth}
\noindent{Algorithm {\sf makeSeekWitnessList}($N_w,d,u,v$)}
\begin{enumerate}
\item Perform a BFS for $d$ levels starting from $u$. Put all vertices
  encountered in $S_d(u)$.
\item Similarly construct $S_d(v)$ by performing a BFS for $d$ levels
  from $v$. 
\item Let $S_w(u,v) \leftarrow S_d(u) \cap  S_d(v)$.
\item For each $x \in S_w(u,v)$ compute $\min\{h(v,x),h(u,x)\}$ where
  $h(x,y)$ is the number of hops in a directed path from $x$ to $y$. 
\item\label{step:sort} Return $S_w(u,v)$ sorted in ascending order of
  $\min\{h(v,x),h(u,x)\}$, breaking ties between two witnesses $x$ and $y$
  by comparing the value of $\max\{h(v,x),h(u,x)\}$ with
  $\max\{h(v,y),h(u,y)\}$, and further breaking ties arbitrarily if
  they are the same.
\end{enumerate}
\end{minipage}
}
\caption{Finding {\em SeekRel} witnesses for $u$ and $v$.}
\label{fig:algorithm:witnesses}
\end{center}
\end{figure}

\noindent{\bf Finding witnesses.} For both {\em SeekRel} and {\em
FactRel} we have to find witnesses in each $N_w$. In
Figure~\ref{fig:algorithm:witnesses} we describe a simple algorithm
that uses breadth-first search from both $u$ and $v$ upto $d$ levels
for some value of $d$ to return a sorted list, $S_w$, of witnesses for
{\em SeekRel}. Note that we do not just create a set of witnesses, but
actually make an ordered list of witnesses. The significance of this
will become clear shortly. In order to construct a list of witnesses
for {\em FactRel} we simply reverse the direction of all the edges of
$N_w$ and execute exactly the same algorithm. We denote the set of
witnesses for {\em FactRel} by $F_w(u,v)$. Note that the reversal of
edges is only to find witnesses, not to compute flows, that takes
place on the same graph for both {\em SeekRel} and {\em FactRel}.

\noindent{\bf Computing flows.} Having constructed the ordered list of
witnesses we go down the list one vertex at a time using any standard
single-source maximum flow algorithm for computing the max flow first
from $u$ to or from the witness as required, then from $v$. Note that
when computing the flow from $u$ to a witness we eliminate $v$ from
the network and vice versa. This is to ensure that the one page
doesn't piggyback on the other in order to route flow to a witness
i.e. all the flow $v$ sends to a witness is independent of $u$ and
vice versa. After computing the flow for a particular witness, we
reduce the capacity of the edges into or out of the witness. Let us
postpone discussing the rationale and method for this to the end of
this section. We then move on to the next witness in the list. See
Figure~\ref{fig:algorithm:seekFlow} for a formal description of the
algorithm for computing flows for {\em SeekRel}. The algorithm for
computing flows for {\em FactRel} is symmetric. In this case we denote
the flow from a witness $x$ as $\factFlow_w(x)$.

\begin{figure}[ht]
\begin{center}
\fbox{\small
\begin{minipage}{0.95\columnwidth}
\noindent{Algorithm {\sf flowSeek}($N_w, d, u,v$)}
\begin{enumerate}
\item $L \leftarrow \mbox{\sf makeSeekWitnessList}(N_w,d,u,v)$.
\item While $L \ne \emptyset$
\begin{enumerate}
\item Extract the first element $x$ from $L$.
\item $\flow_w(u,x) \leftarrow$ max flow from $u$ to $x$ in $N_w
  \setminus \{v\}$.
\item $\flow_w(v,x) \leftarrow$ max flow from $v$ to $x$ in $N_w
  \setminus \{u\}$.
\item $\seekFlow_w(x) \leftarrow \min\{\flow_w(u,x),
  \flow_w(v,x)\}.$
\item\label{step:reduce} Call {\sf reduceSeekCapacity}$(x)$.
\item $L \leftarrow L \setminus \{x\}$.
\end{enumerate}
\end{enumerate}
\end{minipage}
}
\caption{Finding the witness flow for {\em SeekRel}}
\label{fig:algorithm:seekFlow}
\end{center}
\end{figure}

\noindent{\bf The scores.} In order to compute the relationship scores
we have to be able to combine flows from different subnetworks. In
order to do this we normalize all flows by dividing by the weight of
the maximum edge in the network, a quantity we denote by
$\maxwt(w)$. Also, we factor in the relative importance of the various
keywords in $K$ using the function $\gamma$. Hence our first two
scores are
\begin{center}
\fbox{
$\displaystyle \SeekRel(u,v) = \sum_{w\in K}
  \frac{\gamma(w)}{\maxwt(w)} \cdot\left( \sum_{x\in S_w(u,v)}
  \seekFlow_w(x)\right).$} 
\end{center}
\begin{center}
\fbox{
$\displaystyle \FactRel(u,v) = \sum_{w\in K}
  \frac{\gamma(w)}{\maxwt(w)} \cdot \left( \sum_{x\in F_w(u,v)}
  \factFlow_w(x)\right).$} 
\end{center}
The rationale behind {\em FactRel} is that pages providing similar
information will be identified by witnesses in the network, which act
collaboratively to identify good sources of information and link to
them. Similarly, the intuition between {\em SeekRel} is that two pages
which allow a user to reach more or less the same pages are related in
terms of their ability to guide the user as she navigates the Web.

The third measure, {\em SurfRel} is easier to compute since it
requires no witnesses. We simply compute the max flow from $u$ to $v$
in $N_w$, denoting it $\flow_w(u \rightarrow v)$ and the flow from $v$
to $u$, denoted $\flow_w(v \rightarrow u)$. Now, we can say that
\begin{center}
\fbox{
$\displaystyle \SurfRel(u \rightarrow v) = \sum_{w\in K}
  \frac{\gamma(w)}{\maxwt(w)} \cdot \flow_w(u \rightarrow v).$} 
\end{center}
{\em SurfRel} encapsulates the idea that if the Web allows one page to
reach another through a simple sequence of clicks, these two pages
must be related because they are both likely to be visited in a single
browsing session. This idea of two pages being related by the presence
of a path between them is very intuitive and the concept of flow
generalizes the idea of paths. By capacitating edges with the hub
value of the nodes they originate at, we differentiate between nodes
and their ability to allow information to propagate by looking at
their credibility as hubs in the hyperlinked environment. The more the
credibility of the node as a hub, the more the flow it can forward.

\begin{figure}[htbp]
\begin{center}
\epsfig{file=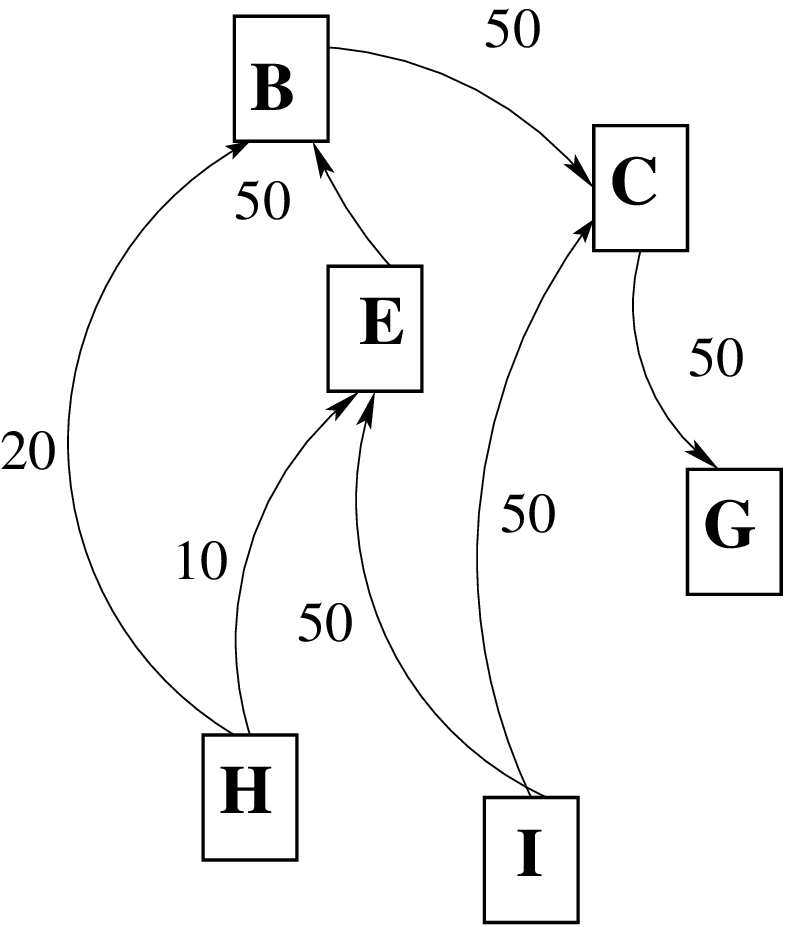,height=2in}
\caption{Redundant witnesses. Capacities are marked on edges.}
\label{fig:redundant-witness}
\end{center}
\end{figure}

\noindent{\bf Reducing witness capacity.} In the case of {\em SeekRel}
and {\em FactRel}, the presence of many witnesses that can sink or
source a lot of flow from the two pages being scored (or send a lot of
flow to them) leads to a higher score for the pair. But there are
cases where this score may be artificially high. Consider the network
in Figure~\ref{fig:redundant-witness}. $E$, $B$, $C$ and $G$ all
witness {\em SeekRel} for $H$ and $I$. But the flow to $B$, $C$ and
$G$ all goes through $E$. So these three are redundant, in the sense
that the information they provide is already contained in the fact
that $E$ is a witness for $H$ and $I$.

\begin{figure}[ht]
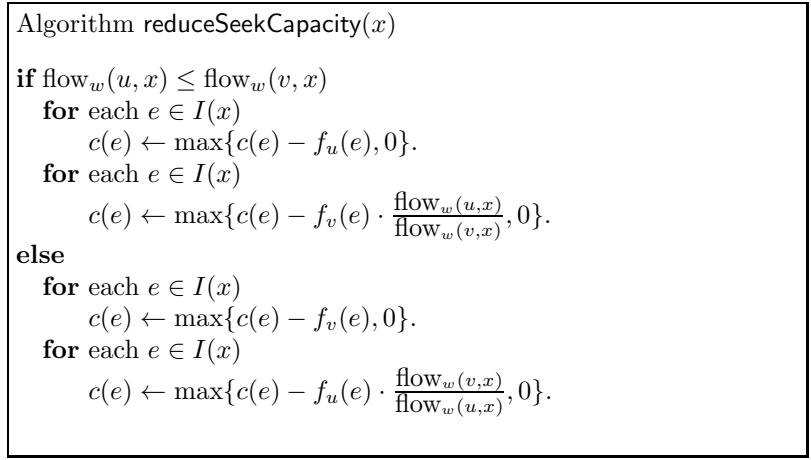

\begin{center}
\fbox{\small
\begin{minipage}{0.6\columnwidth}
\noindent{Algorithm {\sf reduceSeekCapacity}($x$)}
\begin{tabbing}
{\bf if} \= $\flow_w(u,x) \leq \flow_w(v,x)$\\
\> {\bf for} \= each $e \in I(x)$\\
\> \> $c(e) \leftarrow \max\{c(e) - f_u(e),0\}.$\\
\> {\bf for} each $e \in I(x)$\\
\>\> $c(e) \leftarrow \max\{c(e) -   f_v(e)\cdot
\frac{\flow_w(u,x)}{\flow_w(v,x)}, 0\}.$\\
{\bf else}\\
\> {\bf for} each $e \in I(x)$\\
\> \>$c(e) \leftarrow \max\{c(e) -   f_v(e),0\}.$\\
\> {\bf for} each $e \in I(x)$\\
\> \> $c(e) \leftarrow \max\{c(e) -  f_u(e)\cdot
\frac{\flow_w(v,x)}{\flow_w(u,x)},0\}.$\\ 
\end{tabbing}
\end{minipage}
}
\caption{Reducing capacities for {\em SeekRel}}
\label{fig:algorithm:reduce}
\end{center}
\end{figure}

It is to prevent these redundant witnesses from artificially inflating
the relationship score that we reduce the capacity associated with the
witness in Step~\ref{step:reduce} of the flow computing algorithm of
Figure~\ref{fig:algorithm:seekFlow}. For {\em SeekRel} when we are
done computing flow to a witness we reduce its incoming capacity
before moving on to the next witness in the list. For {\em FactRel}
the outgoing capacity is reduced. Before we describe the algorithm
formally in Figure~\ref{fig:algorithm:reduce} let us define some
notation. For a vertex $x$ let the set of incoming edges be $I(x)$ and
the set of outgoing edges be $O(x)$. Let the flow routed for vertex
$u$ on edge $e$ be $f_u(e)$. The capacity of edge $e$ is $c(e)$.

Essentially what {\sf reduceSeekCapacity}($x$) does is remove the
amount of flow witnessed at $x$. Since we take the minimum of
$\flow_w(u,x)$ and $\flow_w(v,x)$ as the amount of flow being
witnessed, we remove this amount from the incoming capacity of
$x$. And to ensure we do this fairly for both $u$ and $v$, we penalize
the incoming edges used by both the flows $\flow_w(u,x)$ and
$\flow_w(v,x)$ equally by scaling down the larger flow to the smaller
one before subtracting it from the capacity of the incoming edge. The algorithm {\sf
reduceFlowCapacity}($x$) is symmetric to this, only it removes
capacity from the outgoing edges of the witness. In both
these cases, with the witness capacity reduced, the ability of
redundant witnesses to skew the flow reduces.

The reason for fixing an ordering for our witnesses becomes clear now
since the capacities of the network decrease after each flow
calculation. Clearly the order in which the witnesses are processed
will make a difference to the flow that is routed to them. Recall the
function $h(u,v)$ defined in Figure~\ref{fig:algorithm:witnesses} as
the number of hops in a directed path from $u$ to $v$. Let us see what
the values of $(h(x,v), h(x,u))$ are for $v = H$ and $u = I$. For $G$
this is (3,2), for $C$ it is (2,1), for $B$ it is (1,2) and for $E$ it
is (1,1). So the sorted order according to our algorithm should be
$E$, $B$, $C$, $G$. For this the total flow witnessed will be 10 units
for $E$ plus 20 units for $B$ = 30 units. Because of capacity
reduction $C$ and $G$ will not be able to witness any flow. But is
this correct given that $I$ has a direct edge to $C$ which doesn't go
through $B$ or $E$? We argue it is since the flow of 30 that could be
witnessed at $C$ is the same as the flow of 30 units that is being
witnessed for $H$ at $B$ and $E$. In
Figure~\ref{fig:redundant-witness} for example, if we choose the
witnesses in the reverse order $G, C, B, E$ the total flow is 30
witnessed by $E$ + 30 witnessed by $C$ + 30 witnessed by $B$ + 10
witnessed by $E$, totalling 100, of which 70 units are redundant.

Thus it follows that distant witnesses along a chain of nodes are more
likely to be redundant since decreasing their capacity will not affect
the flow to nearer witnesses. With this in mind, and also noting that
more information about a Web page is likely to be found in its near
neighborhood rather than far away from it, we order witnesses in
increasing order of the distance of the witness from one of the pages
in (Step~\ref{step:sort} of algorithm {\sf
makeSeekWitnessList}($N_w,d,u,v$) in
Figure~\ref{fig:algorithm:witnesses}.  If a distant witness still
witnesses flow after the reduction of capacity of nearer witnesses, we
can be sure that this is not redundant flow.

\subsection{Discussion}
\label{sec:algorithm:discuss}

\begin{figure}[htbp]
\begin{center}
\epsfig{file=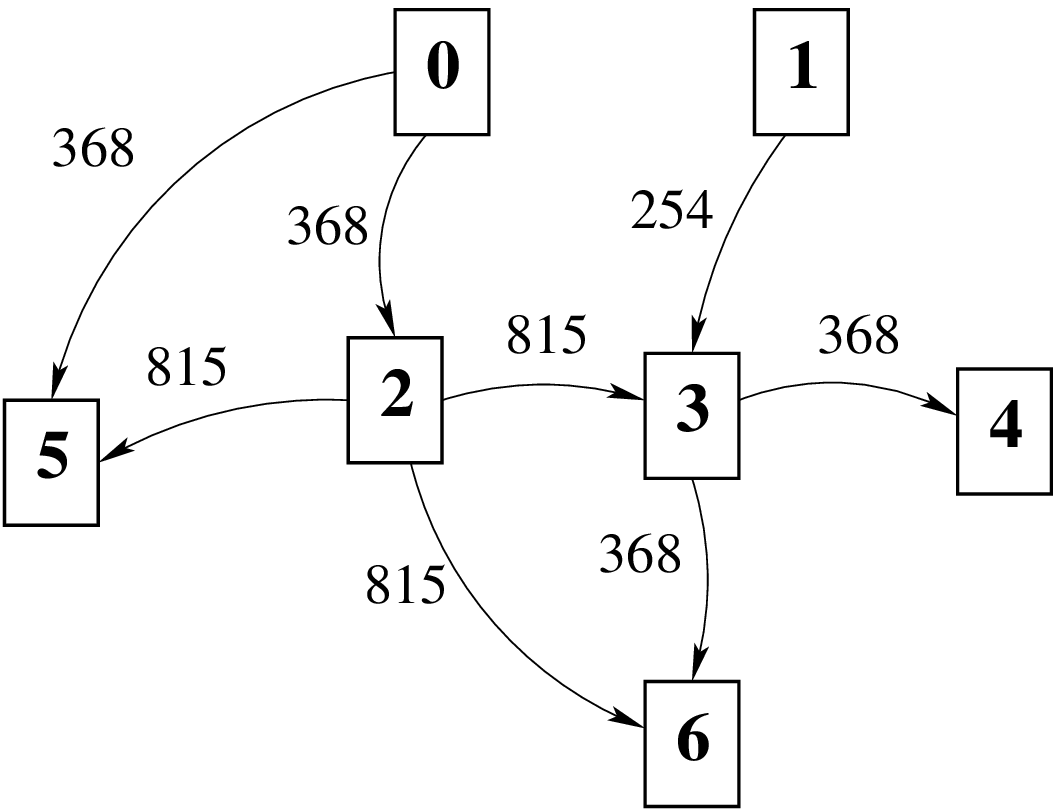,width=0.5\columnwidth}
\caption{A simple subnetwork with edge capacities according to the hub
values of the originating node.}
\label{fig:simrank-example}
\end{center}
\end{figure}

We took the simple subnetwork of Figure~\ref{fig:simrank-example} and ran
our scoring algorithms on it. The table of scores obtained is in
Figure~\ref{fig:toy-table}. For cleanness of presentation all hub
values have been scaled by 1000. The flow values have been scaled up
by $\maxwt = 815$ since we are only considering one subnetwork.

\begin{figure*}[ht]
\begin{center}
{\tiny
\begin{tabular}{|c||c|c|c|c|c|c|c|}
\hline
  & 0&	1&	2&	3&	4&	5 & 6 \\
 \hline\hline
0 &  & (253,0,0,0) & (368,0,368,0) & (368,0,368,0) &
(0,0,368,0) & (0,0,736,0) & (0,0,368,0)\\
\hline
1 &
(253,0,0,0)&&(253,0,0,0)&(0,0,253,0)&(0,0,253,0)&(0,0,0,0)&(0,0,253,0)\\
\hline
2 &
(368,0,0,368)&(253,0,0,0)&&(368,0,815,0)&(0,0,368,0)&(0,368,815,0)&(0,0,1183,0)
\\
\hline
3 &
(368,0,0,368)&(0,0,0,253)&(368,0,0,815)&&(0,0,368,0)&(0,815,0,0)&(0,815,368,0)
\\
\hline
4 &
(0,0,0,368)&(0,0,0,253)&(0,0,0,368)&(0,0,0,368)&&(0,736,0,0)&(0,368,0,0)	\\
\hline
5 &
(0,0,0,736)&(0,0,0,0)&(0,368,0,815)&(0,815,0,0)&(0,736,0,0)&&(0,1183,0,0)
\\
\hline
6 &
(0,0,0,368)&(0,0,0,253)&(0,0,0,1183)&(0,815,0,368)&(0,368,0,0)&(0,1183,0,0)&	\\
\hline
\end{tabular}
\\
Legend: ($\SeekRel(x,y), \FactRel(x,y), \SurfRel(x \rightarrow y), \SurfRel(y \rightarrow x)$)
}
\caption{Relationship scores for the network of Figure~\ref{fig:simrank-example} (scaled up by $1000 \times \maxwt$)}
\label{fig:toy-table}
\end{center}
\end{figure*}

\begin{figure}
\begin{center}
{\small
\begin{tabular}{|c||c|c|c|c|}
\hline
&  $\SeekRel$ & $\FactRel$ & $\SurfRel \rightarrow$ & $\SurfRel
\leftarrow$\\
\hline\hline
0&  2,3 &None& 5 &None	\\
\hline
1&  0,2 &None& 3,4,6 &None	\\
\hline
2 &  0,3 & 5 & 6 & 0 	\\
\hline
3 &  0,2 & 5,6 & 4,6 & 2 \\	
\hline
4 & None& 5 &None& 0,2,3 	\\
\hline
5 & None& 6 &None& 2 	\\
\hline
6 & None& 5 &None& 2 	\\
\hline
\end{tabular}
}
\caption{High scorers for Figure~\ref{fig:simrank-example} }
\label{fig:toy-relations}
\end{center}
\end{figure}

In Figure~\ref{fig:simrank-example} we see that node 2 is related by
{\em SeekRel} to nodes 0 and 3. This is as it should be since they
both point to similar parts of the network. Node 0 sends 368 units of
flow to the witness 5 which is matched by 2. Node 3 sends 368 units of
flow to witness 6, which is also matched by 2. And although the node 1
shares many witnesses with 0, the flow it can send is limited by its
outgoing capacity (which is low because it is not a good hub) and so
its {\em SeekRel} score is low, though non-zero, and 2 and 3 beat it
out in scoring.

If we look at the second column of the table in
Figure~\ref{fig:toy-relations}, we see that 5 and 6 are the top
scorers for {\em FactRel} for several nodes. Node 2's high credibility
as a hub helps draw attention to 5 and 6. The node 0's good hub value
helps relate 2 to 5, in what can seen to be an example of pure
co-citation.

In the case of {\em SurfRel}, the example of 0 is interesting. While 0
is strongly related to 5 by this measure, its score with 6 is
relatively less. In this case our score captures the fact that there
are multiple {\em independent} paths from 0 to 5 while all paths from
0 to 6 go through 2. The divergence in the paths after 2 does not help
boost {\em SurfRel} for 0 and 6.

To further illustrate the power of our methods, we implemented the
{\em SimRank}~\cite{simrank} and {\em PageSim}~\cite{pagesim} scoring
algorithms and scored our simple subnetwork using them. The
results are in Figures~\ref{fig:simrank-toy-table}
and~\ref{fig:pagesim-toy-table}.

\begin{figure}[ht]
\begin{center}
{\small
\begin{tabular}{|c||c|c|c|c|c|c|c||c|}
\hline
  & 0&	1&	2&	3&	4&	5 & 6 & High Score\\
\hline\hline
0&1.0&0.0&0.0&0.0&0.0&0.0&0.0&None\\
\hline
1&0.0&1.0&0.0&0.0&0.0&0.0&0.0&None\\
\hline
2&0.0&0.0&1.0&0.0&0.0&0.5&0.0& 5\\
\hline
3&0.0&0.0&0.0&1.0&0.0&0.25&0.25& 5,6\\
\hline
4&0.0&0.0&0.0&0.0&1.0&0.0&0.5& 6\\
\hline
5&0.0&0.0&0.5&0.25&0.0&1.0&0.25& 2 \\
\hline
6&0.0&0.0&0.0&0.25&0.5&0.25&1.0& 4\\
\hline
\end{tabular}
}
\caption{{\em SimRank} scores for Figure~\ref{fig:simrank-example}.}
\label{fig:simrank-toy-table}
\end{center}
\end{figure}

{\em SimRank} related none of the pages to either 0 or 1 whereas our {\em
SeekRel} is able to detect the fact that 0 can aid in helping the user
find links to pages that 2 and 3 can also lead to. Even 1 shares this
property as a navigational aid with some of the other pages, a fact
that comes up in our scoring. The case of node 1 is particularly
interesting because {\em PageSim}, that gives non-zero scores in many cases
where {\em SimRank} fails, does not deduce 1's relationship to 0 and 2 that
our {\em SeekRel} is able to find. A user currently viewing 1 would
come to believe that only 3 and 4 are related to 1 if she relied on
{\em PageSim} or {\em SimRank}. This would be erroneous because a knowledge that 0
is related in terms of links it provides could lead that user to 2,
which she would not find if she relied on these other two measures.
{\em PageSim} is somewhat more sophisticated than {\em SimRank} so it detects 0's
relationship to 5, just like our $\SurfRel \rightarrow$ does, but it
misses 0's relationship to 3 that we find through {\em SeekRel}.

\begin{figure}[ht]
\begin{center}
{\small
\begin{tabular}{|c||c|c|c|c|c|c|c||c|}
\hline
  & 0&	1&	2&	3&	4&	5 & 6 & High Score\\
\hline
0&0.08&0.0&0.04&0.01&0.01&0.05&0.02& 5 \\
\hline
1&0.0&0.08&0.0&0.08&0.04&0.0&0.04& 3 \\
\hline
2&0.04&0.0&0.16&0.05&0.03&0.08&0.08& 5,6 \\
\hline
3&0.01&0.08&0.05&0.33&0.16&0.05&0.19& 6 \\
\hline
4&0.01&0.04&0.03&0.16&0.33&0.03&0.16& 3,6 \\
\hline
5&0.05&0.0&0.08&0.05&0.03&0.25&0.06& 2 \\
\hline
6&0.02&0.04&0.08&0.19&0.16&0.06&0.42& 3 \\
\hline
\end{tabular}
}
\caption{{\em PageSim} scores for Figure~\ref{fig:simrank-example}.}
\label{fig:pagesim-toy-table}
\end{center}
\end{figure}

{\em PageSim} almost misses 5's relationship to 4 and also scores 5's
relationship to 6 quite low. {\em SimRank} completely misses the
relationship to 4 and scores the relationship to 6 lower than the
relationship to 2. On the other hand, a high {\em FactRel} score for
both of these allows a user to tell that the information available at
4 and 6 are both relevant to people who are interested in 5. Since our
{\em FactRel} score between 5 and 2 is relatively lower and our {\em
  SurfRel} score between them is high, a user can deduce the nature of the relationship between 5 and 2, a fact also detected by {\em SimRank}.

We now move on to experiments on real data taken from the Web.

\section{Experimental evaluation}
\label{sec:experiments}

\subsection{Experimental setup}
\label{sec:experiments:setup}

We performed our experiments on four data sets taken from the
Web. Creating these data sets was a multi-stage process that began by
querying AltaVista~\cite{altavista} with a search string and taking
the top 100 results to form a core set. We did not use Google since we
compare our results to Google's Similar Pages feature. We then used
the open source Web crawler Nutch~\cite{nutch} to retrieve the pages
linked from the core set. Then we found the top 1000 pages that link
to these new pages using Altavista's advanced feature providing
inlinks for a queried page. Finally, we found the inlinks of the pages
in the core using Altavista then went back to Nutch to find the
outlinks of these pages. We followed Dean and Henzinger~\cite{dean}
and took only the top 10 outlinks in the manner they specified i.e. if
we were looking at the outlinks of a page $u$ which pointed to a core
page $v$, we took only the links on $u$ which were ``around'' the link
to $v$ in the sense that we took the 5 links immediately preceding the
link to $v$ on the page and the 5 links immediately following
$v$. Having obtained this data set we preprocessed it by computing the
hub and authority values of all the pages in it.

Our four data sets were generated using the keyword strings
``automobile'' (54952 pages), ``motor company'' (14973 pages),
``clothes shopping'' (37724 pages) and ``guess'' (12101
urls). For repeatability purposes, these data sets have been
made available online.\footnote{http://www.cse.iitd.ernet.in/$\tilde{\
}$bagchi/relationship-scores/}  We conducted extensive experiments on all these data sets by taking one page out of them as a query, then
scoring all three relationships for this page with all the other pages
in the data set. Note that this does not exactly correspond to our
claim in Section~\ref{sec:algorithm} that we proceed by extracting
keywords from pairs of pages and then computing flows on the
subnetworks obtained from those keywords. Limitations on the amount of
data we were equipped to handle in a university setting prevented us
from performing these steps in the full. We present here these
stripped down experiments as indicative of what a full implementation
of our scoring mechanisms might be able to achieve.

We compared our 10 top scoring pages for {\em FactRel} and {\em
SeekRel} with the top 10 pages returned by Google's Similar Pages
feature. We also implemented the {\em Companion} algorithm described
in~\cite{dean} and compared our results to the top 10 results returned
by it. For {\em SurfRel} we simply took our top 10 results and
evaluated them. The evaluation in all these cases was done by
conducting user surveys.

For each of the target URLs scored using {\em FactRel} we asked the
user to imagine they had visted it in the course of an
information-gathering task and found it relevant. We then assembled a
set of 30 URLs ({\em FactRel}'s top 10, Google's top 10 and
{\em Companion}'s top 10). We presented these 30 URLs in a random order and
asked users to answer three yes/no questions: 1) {\em Would you visit
this page if you had already visited the target page?} 2) {\em Does
this page provide similar information to the target page?} and 3) {\em
Is this page relevant to your information-gathering task?}

Each such survey was given to between 5 and 8 users. For each of the
30 pages, and each of these 3 questions a relevance score was
computed. For a given target URL $t$, and a result page $i$:
\[\rel(t,i) = \frac{\mbox{Number of YES answers} }{\mbox{Number of users
who took the survey}}.\]
For each target page and each algorithm, the {\em Precision at $r$} of
the ranked results was computed using the formula
\[\precr(t,r) = \frac{\sum_{i=1}^{r}\rel(i)}{r}.\]
The precision at $r$ for an algorithm was computed by taking the
average of the precision at $r$ values over all the target pages
evaluated. 

For {\em SeekRel} the first and the third question remained the
same. The second question was replaced with {\em Does this page provide
links similar to those in the target page?}. Precision at $r$ was
calculated similarly for {\em SeekRel}. For {\em SurfRel} we only
asked one question: {\em Is this page relevant to your task?} and we
did not present results from Google or {\em Companion}. 

The code for all scoring mechanisms was written in Java
(JDK-1.5.0). The open source Nutch crawler was downloaded and run and
a parser was written in C++ to parse its output. A total of 9 target
pages were evaluated for {\em FactRel}, 7 for {\em SeekRel} and 5 for
{\em SurfRel}. On the 14973 page ``motor company'' data set finding
all three relationship scores between a given query page and all the
other query pages on a desktop PC with a 3.4 GHz Intel Pentium
processor with 1GB RAM took about 8 minutes on average for this data
set.  On the 37,724 page ``clothes shopping'' data set it took about 1
hour on average to calculate all three scores of a given page with all
other pages. Let us now see what the experiments revealed.

\subsection{Experimental results}
\label{sec:experiments:results}

In Figure~\ref{fig:FactRelHonda} we list the 10 URLs that scored the
highest on {\em FactRel} for the page www.honda.com. The relationship
revealed is expected: other major car companies. More interesting is a
list of pages related by {\em FactRel} to www.cngvehicle.com. Not only
do we get pages related to other alternate fuels (Biodiesel Forum
(forums.biodiesel.com), Electric Drive Transportation Association
(evaa.org) and government agencies dealing with renewable energy
policy, we also get links to private car industry players who are
pursuing the development of energy efficient cars.

\begin{figure}[htbp]
\begin{center}
{\small
\begin{tabular}{|c||c||c|}
\hline
 & www.honda.com  & www.cngvehicle.com \\
\hline\hline
1& www.ford.com & www.evaa.org \\
\hline
2& www.toyota.com & forums.biodieselnow.com \\
\hline
3 & www.landrover.com & www.eere.energy.gov/cleancities \\
\hline
4 & www.audi.com & www.ford.com \\	
\hline
5 & www.gm.com & www.nrel.gov \\
\hline
6 & www.mercuryvehicles.com  & www.eere.energy.gov/cleancities/-- \\
\hline
7 & www.cadillac.com  & www.gsa.gov/Portal/gsa/ep/--- \\
\hline
8 & www.chevrolet.com & www.mercuryvehicles.com \\
\hline
9 & www.lincoln.com & www.gm.com \\
\hline
10 & www.porsche.com & www.honda.com\\
\hline
\end{tabular}
}
\caption{{\em FactRel} top scorers for two pages of the ``motor
  company'' data set.}
\label{fig:FactRelHonda}
\end{center}
\end{figure}

For the Web page of the clothing store
Bloomingdales\footnote{www.bloomingdales.com}, {\em FactRel} and
Google's ``similar pages'' returned more or less identical lists right
at the top, but {\em FactRel}'s high scorers remained very focused
(other apparel stores) while Google provided links to coffee shops and
hotels which could possibly be appropriate in some contexts but
deviate from what is arguably the main focus of a user visiting the
Bloomingdale's Web page (see Figure~\ref{fig:FactRelBloomingdales}).

\begin{figure}
\begin{center}
{\small
\begin{tabular}{|c||c||c|}
\hline
 & {\em FactRel} & Google's ``similar pages'' \\
\hline\hline
1 & www.macys.com & www.saksfifthavenue.com\\
\hline
2 & www.neimanmarcus.com & www.macys.com\\
\hline
3 & www.jcpenney.com & www.nordstrom.com\\
\hline
4 & www.abercrombie.com & www.neimanmarcus.com\\
\hline
5 & www.bananarepublic.com & www.barneys.com\\
\hline
6 & www.bluefly.com & www.fds.com\\
\hline
7 & www.spiegel.com & www.starbucks.com\\
\hline
8 & www.saksfifthavenue.com & www.walmart.com\\
\hline
9 & www.target.com & www.nycvisit.com\\
\hline
10 & www.eddiebauer.com & www.ritzcarlton.com\\
\hline
\end{tabular}	
}
\caption{{\em FactRel} vs Google's ``similar pages'' for www.bloomingdales.com} 
\label{fig:FactRelBloomingdales}
\end{center}
\end{figure}

As another demonstration of {\em FactRel}'s reliability in providing
alternate sources of information, we present its top scorers for
www.mysimon.com, a comparison shopping site, in
Figure~\ref{fig:FactRelMysimon}. {\em FactRel} scored Web pages for
major comparison shopping sites very high.

\begin{figure}
\begin{center}
{\small
\begin{tabular}{|c||c|}
\hline
 & {\em FactRel} for www.mysimon.com\\
\hline\hline
1 & www.dealtime.com \\
\hline
2 & shopping.yahoo.com\\
\hline
3 & www.ebay.com\\
\hline
4 & www.bizrate.com\\
\hline
5 & www.pricegrabber.com\\
\hline
6 & www.nextag.com\\
\hline
7 & www.become.com\\
\hline
8 & www.alibris.com\\
\hline
9 & www.buy.com\\
\hline
10 & www.www.bestbuy.com\\
\hline
\end{tabular}	
}
\caption{{\em FactRel} high scorers for www.mysimon.com}
\label{fig:FactRelMysimon}
\end{center}
\end{figure}

\begin{figure}[htbp]
\begin{center}
\epsfig{file=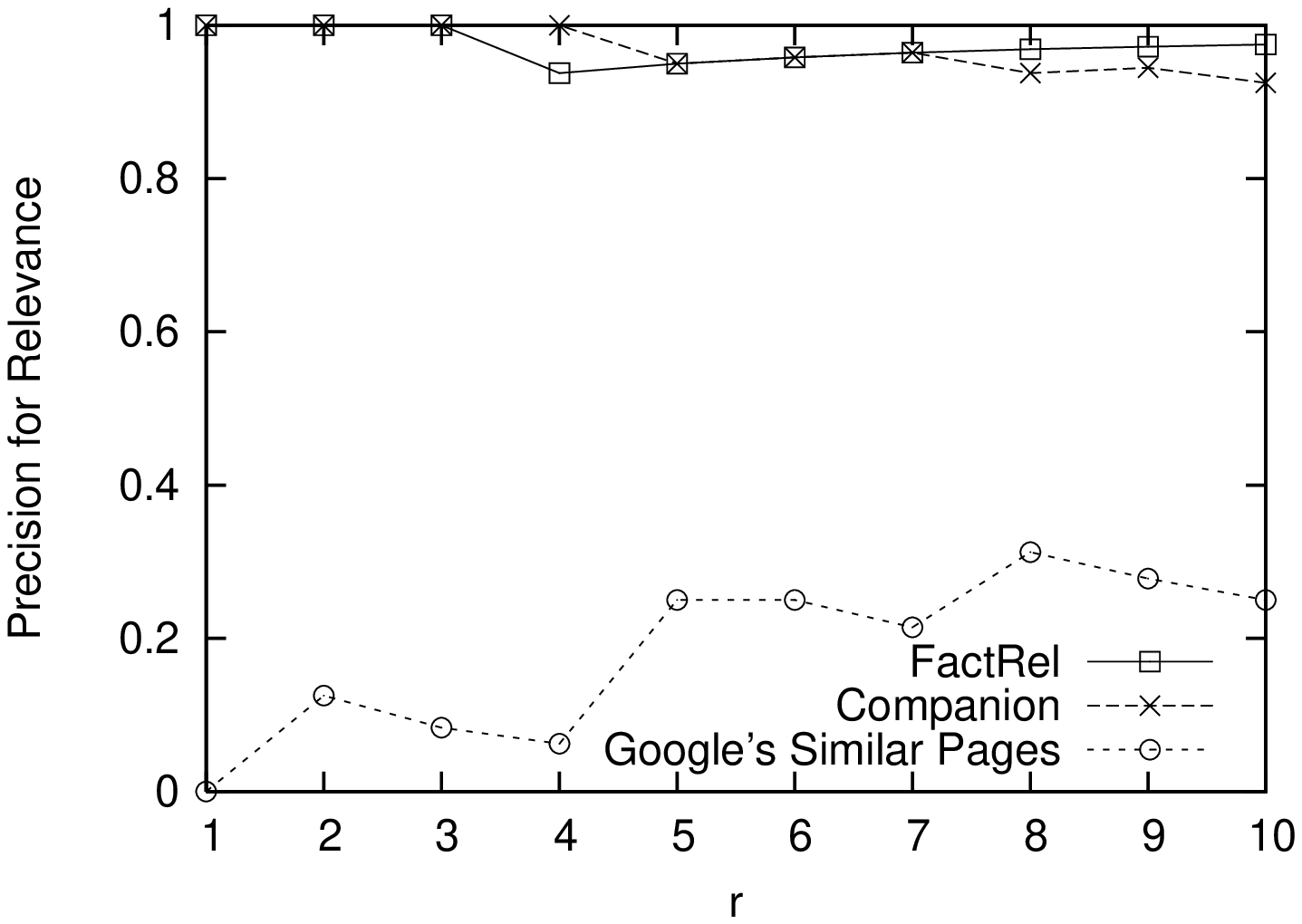,width=0.5\columnwidth,angle=270}
\caption{Precision at $r$ for the relevance question for www.guess.com.}
\label{fig:guess-relevance-precision}
\end{center}
\end{figure}

To test the robustness of {\em FactRel} we scored the home page of
Guess Jeans (www.guess.com) using not just the ``clothes shopping''
data set but also the ``guess'' data set. Despite the presence of the
ambiguous keyword ``guess'', {\em FactRel}'s top 10 results were
closely related to the original page: pages relating to clothing and
accessories. Google's similar pages, on the other hand, appeared to
get severely misled by the keyword ``guess'' (see
Figure~\ref{fig:FactRelGuess}). The precision at $r$ graph for www.guess.com in
Figure~\ref{fig:guess-relevance-precision} reveals that Google does
very poorly while {\em FactRel} and {\em Companion} provide good results.

\begin{figure}
\begin{center}
{\small
\begin{tabular}{|c||c||c|}
\hline
 & {\em FactRel} & Google's ``similar pages''\\
\hline\hline
1 & www.gap.com & www.guessthename.com\\
\hline
2 & www.gucci.com & www.onlineshoes.com/...\\
\hline
3 & www.marciano.com & www.sonypictures.com/...\\
\hline
4 & www.guessinc.com & 	www.imdb.com/title/tt0372237\\
\hline
5 &www.jcrew.com & www.amazon.com/Guess/...\\
\hline
6 & www.gbyguess.com & www.bizrate.com/...guess+bags.html\\
\hline
7 & www.hugoboss.com & www.learner.org/...\\
\hline
8 & www.givenchy.com & popular.ebay.com/...Guess+Jeans.html\\
\hline
9 & www.gianfrancoferre.com & www.answers.com/topic/guess-inc\\
\hline
10 & www.diesel.com & www.guessfinancial.com\\
\hline
\end{tabular}
}
\caption{{\em FactRel} vs Google's ``similar pages'' for
  www.guess.com} 
\label{fig:FactRelGuess}
\end{center}
\end{figure}

\begin{figure}[htbp]
\begin{center}
\epsfig{file=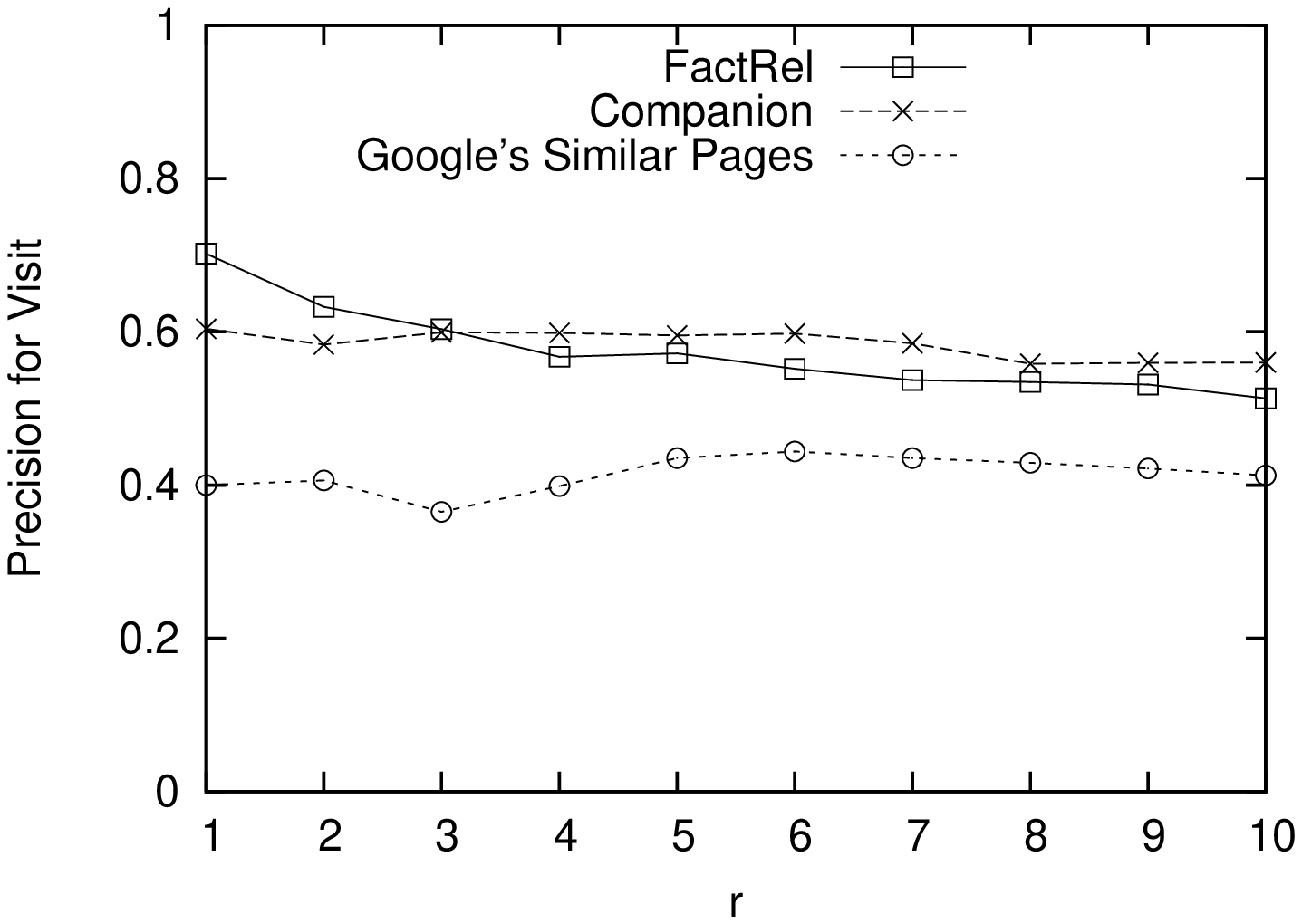,width=0.5\columnwidth,angle=270}
\caption{Precision at $r$ for the visit question for the {\em FactRel}
  target pages.}
\label{fig:auth-visit-precision}
\end{center}
\end{figure}

In general we found that {\em FactRel}'s results were substantially
better than Google's but not better than those returned by
{\em Companion}. In Figure~\ref{fig:auth-visit-precision} we see that users
preferred to visit the top 10 pages presented by {\em FactRel} over
those of Google after having visited the target page.

\begin{figure}[htbp]
\begin{center}
\epsfig{file=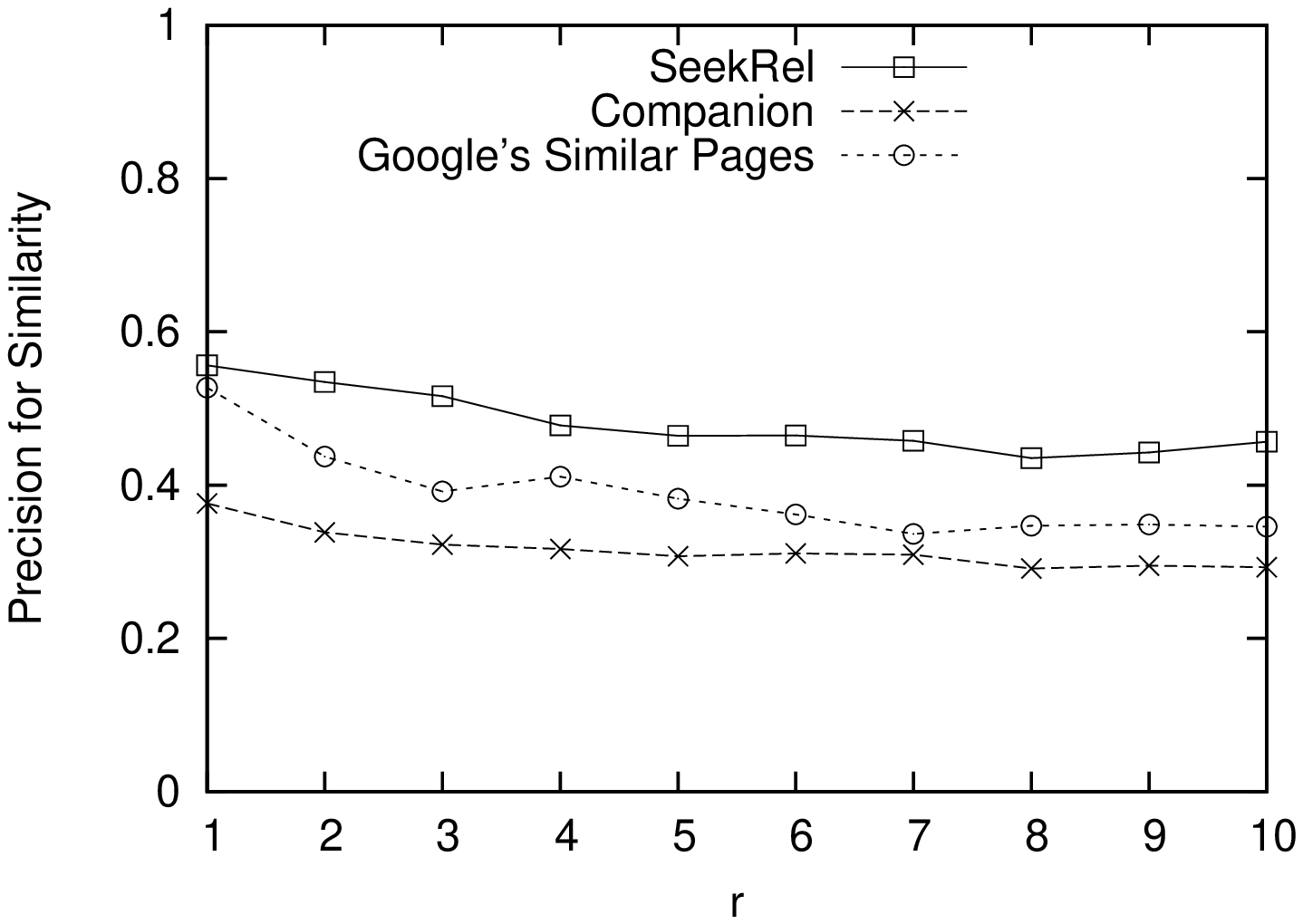,width=0.5\columnwidth,angle=270}
\caption{Precision at $r$ for the similar links question for the {\em SeekRel}
  target pages.}
\label{fig:hub-similar-precision}
\end{center}
\end{figure}

\begin{figure}[htbp]
\begin{center}
\epsfig{file=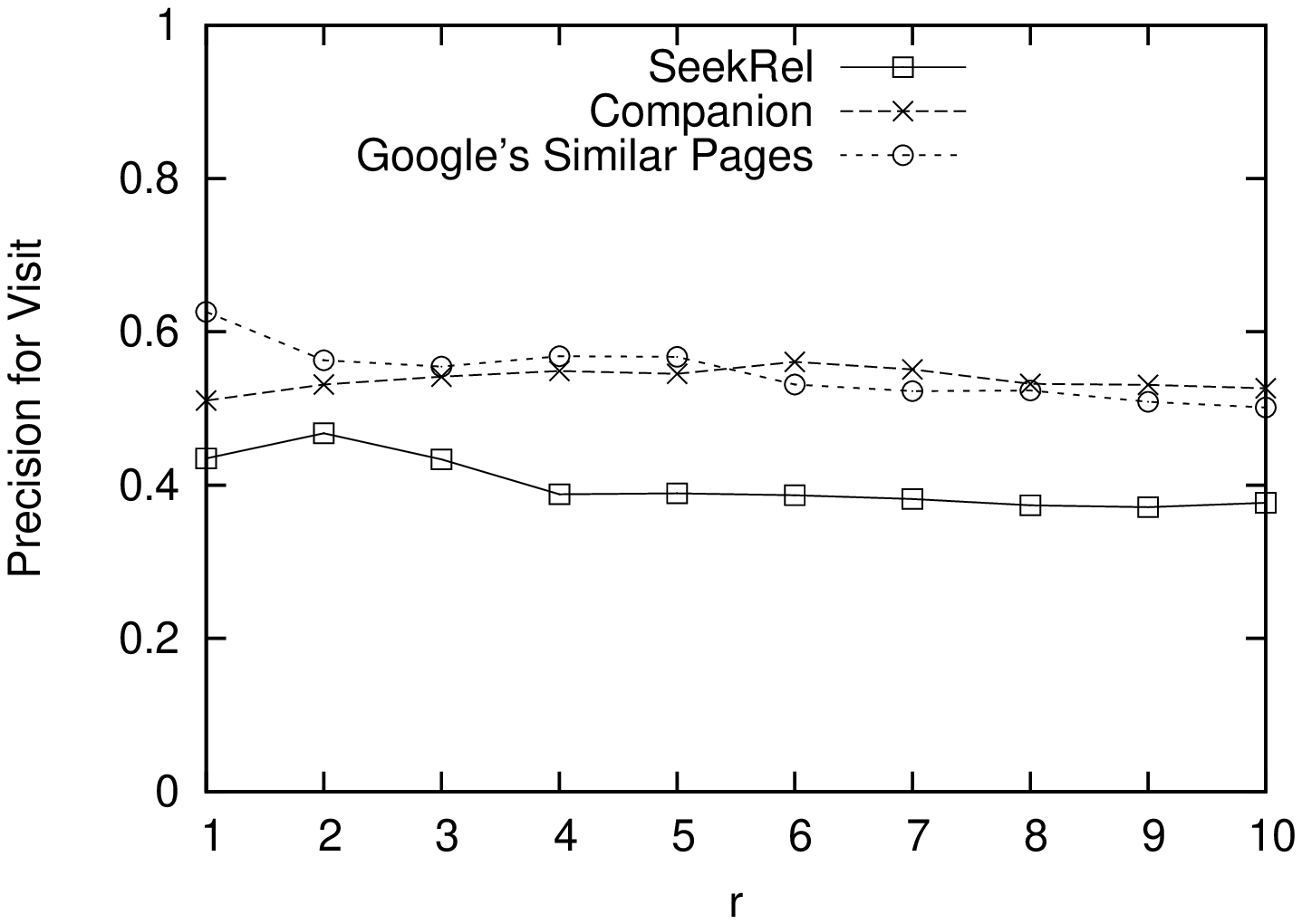,width=0.5\columnwidth,angle=270}
\caption{Precision at $r$ for the visit question for the {\em SeekRel}
  target pages.}
\label{fig:hub-visit-precision}
\end{center}
\end{figure}

For {\em SeekRel} the picture was more complex. While users generally
felt that the pages presented by {\em SeekRel} were far better than
the results presented by Google and {\em Companion} in terms of the
similarity of the links on them to those on the target page (see
Figure~\ref{fig:hub-similar-precision}), they preferred to vist the
pages presented by the other two algorithms (see
Figure~\ref{fig:hub-visit-precision}). This is strong independent
evidence in favour of Aula et. al.'s conclusion~\cite{aula:2005} that
Web users prefer to browse rather than search. Rather than visit
another page with links similar to a given page they would rather
visit a page with actual information on it.

Another possible drawback in {\em SeekRel} was revealed when we scored
for a page listing all the malls in the Santa Clara, California
area\footnote{www.ersys.com/usa/06/0669084/mall.htm.} (see
Figure~\ref{fig:SeekRelErsys}). We found that
pages with links to online shopping resources and even personal
shopping options scored high. But we also found that pages with local
information for places as far afield from Santa Clara as Macon,
Georgia appeared near the top of the list. The absence of geographical
domain knowledge in our system shows up here.

\begin{figure}
\begin{center}
{\small
\begin{tabular}{|c|}
\hline
www.shoppingcolumn.com/personal-shopping.html \\
\hline
www.misslist.com/stores/shoes.html \\
\hline
www.arlingtoncards.com/aroundtown/bizshop1.htm \\
\hline
www.allwi.com/wipresents.html \\
\hline
www.digital-librarian.com/shopping.html \\
\hline
www.cool1055.com/lc/features/shopping \\
\hline
www.ersys.com/usa/13/1349000/mall.htm \\
\hline
\end{tabular}
}
\caption{{\em SeekRel} top scorers for ersys.com's mall information
  page for Santa Clara, CA.}
\label{fig:SeekRelErsys}
\end{center}
\end{figure}

\begin{figure}[htbp]
\begin{center}
\epsfig{file=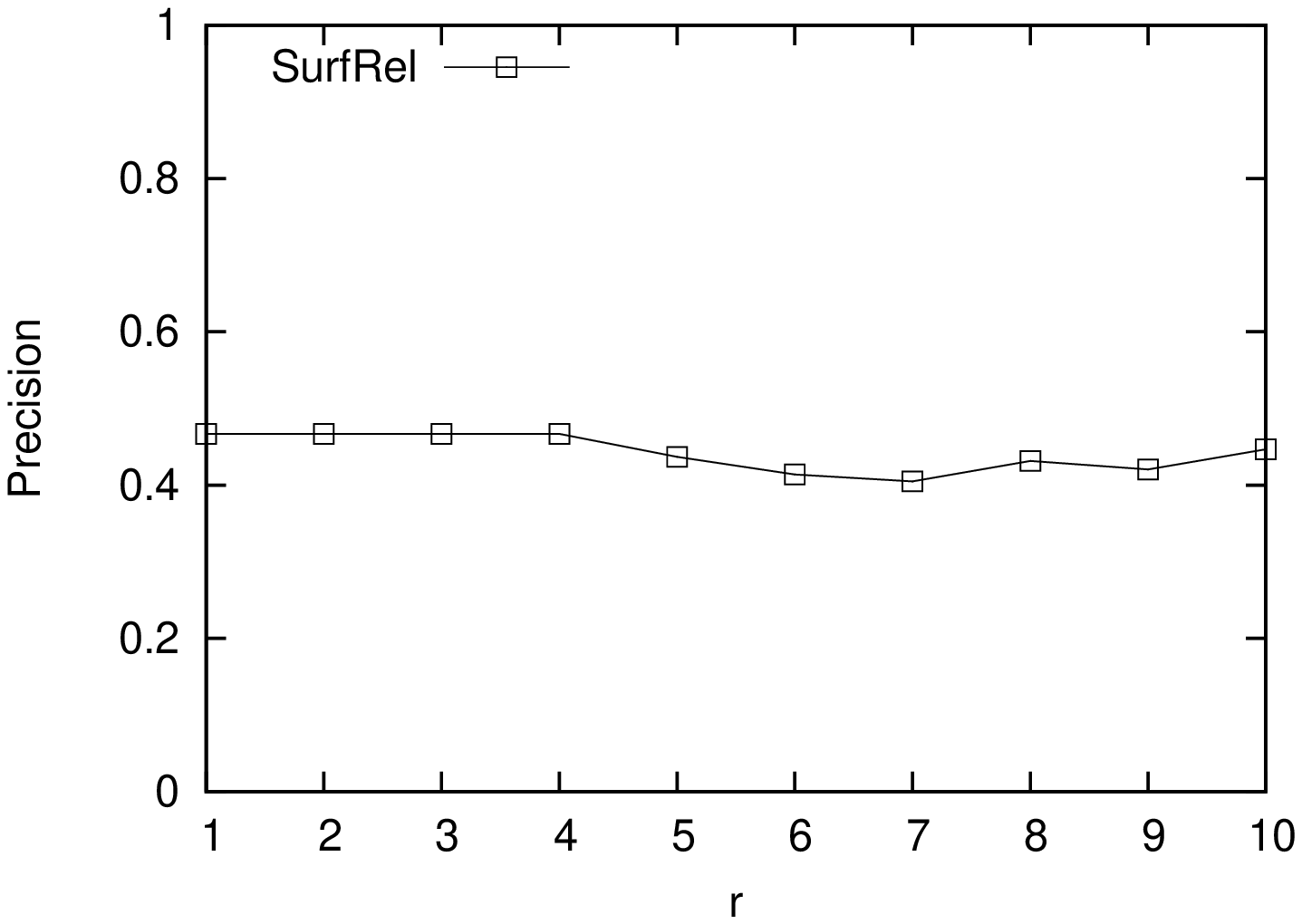,width=0.5\columnwidth,angle=270}
\caption{Precision at $r$ for {\em SurfRel}.}
\label{fig:surf-relevance-precision}
\end{center}
\end{figure}

The user response to {\em SurfRel} was fairly good. Almost half of our
top ten results were found relevant by the respondents (see
Figure~\ref{fig:surf-relevance-precision}).

\section{Discussion} 
\label{sec:discussion}

Our scoring mechanisms are designed with a view to integrate the two
broad streams of thought on web page relationships: textual and
link-based. This goes some way in addressing Lawrence and Giles'
criticism that search engines are biased towards pages which are
well-linked~\cite{lawrence:1999} and is hence an advantage over
algorithms like Dean and Henzinger's {\em Companion} which take only link
information into consideration.

Although users ranked our results in the same ballpark as {\em
Companion}, it is our contention that our algorithms are much less
resource intensive and much more suited to inclusion in a real-world
search engine. We maintain a deck of subnetworks, one corresponding to
each significant keyword. It is difficult to estimate the number of
significant keywords but if we take the widely used lexical database
WordNet~\cite{wordnet} as an indicator, the number is of the order of
100,000.\footnote{WordNet 3.0 contains 155,287 distinct strings.} The
{\em Companion} algorithm, on the other hand, creates a subnetwork for
each queried page. Creating the subnetwork at query time is difficult
because of the overhead involved in crawling the Web (or even an image
of the web stored on disk) and preprocessing and storing these
networks appears infeasible given that the size of the World Wide Web
is estimated to be in the tens of billions of
pages~\cite{deKunder:2008}. Even if it were feasible to store
structural information on such a scale, the problem of updation is
hard to solve. The Web is constantly changing, and updating our
keyword-based subnetworks will be an order of magnitude less resource
intensive than updating billions of page-specific subnetworks.

Our approach is further vindicated by the observation that the results
provided by {\em FactRel} and {\em SeekRel} clearly outperform
Google's Similar Pages. We have tried to gracefully bring together
textual and link information in a common framework where one can
compensate for the shortcomings of the other.  The Guess Jeans example
presented above demonstrates that our scores can leverage link
information to handle ambiguous keywords in a manner better than
Google can.

The main contribution of this paper, in our view, is the location of
our thinking on how to relate pages in the context of user intent. As
part of our future research agenda we want to formulate relationships
between pages that can service user intent outside the domain of
information-gathering. We also want to test the applicability of our
methods in social networking situations and user-generated content
scenarios.

\ignore{
\section{Conclusion}
\label{sec:conclusion}

In this paper we have brought the notion of user intent to bear on the
relationships between Web pages. We have particularly focused on
information-gathering tasks and pointed out how the different stages
of an information-gathering task can be enabled by providing the user
with pages that bear a relationship to the page currently being viewed
either in terms of the real world information they provide or the
navigational information they provide.  As part of our future research
agenda we want to study user intent outside the domain of
information-gathering, and try to formulate relationships between
pages that can service that intent. We also want to test the
applicability of our methods in social networking situations and
user-generated content scenarios. The notion of mashups and dynamic
content is increasingly making the view of the Web as a set of static
pages linking to each other irrelevant. When the notion of Web page
itself is being transformed, the relationship between pages will also
be transformed. Understanding this transformation, contextualizing the
new relationships and developing methods to quantify them will be an
important challenge going into the future.  }

\end{document}